# RobustNLP: A Technique to Defend NLP Models Against Backdoor Attacks

MARWAN OMAR

Illinois Institute of Technology

Abstract

As machine learning (ML) systems are being increasingly employed in the real world to handle sensitive tasks and make decisions in various fields, the security and privacy of those models have also become increasingly critical. In particular, Deep Neural Networks (DNN) have been shown to be vulnerable to backdoor attacks whereby adversaries have access to the training data and the opportunity to manipulate such data by inserting carefully developed samples into the training dataset. Although the NLP community has produced several studies on generating backdoor attacks proving the vulnerable state of language modes, to the best of our knowledge, there does not exist any work to combat such attacks. To bridge this gap, we present RobustEncoder: a novel clustering-based technique for detecting and removing backdoor attacks in the text domain. Extensive empirical results demonstrate the effectiveness of our technique in detecting and removing backdoor triggers. Our code is available at https://github.com/marwanomar1/Backdoor-Learning-for-NLP

## 1 INTRODUCTION

The unique capabilities offered by Machine Learning (ML) models in automating tasks and solving complex business challenges have led to the widespread adoption of those models in various fields and applications, including business, finance, healthcare, and education, to name a few. However, the same ML models and the systems that utilize them are not developed in a fashion that ensures secure operation when deployed to production networks in the real world. Although the performance of ML models is measured during the training time using the test dataset, this is not sufficient because it does not factor in the risk of malicious attacks on such systems after deployment. An extensive body of research exists on defending against adversarial examples in the image and NLP domains. In the NLP context, adversarial examples work by perturbing the input text to fool a classifier into producing incorrect results (i.e., misclassification of input text) [4]. However, because the training process is not always fully controlled by business entities deploying NLP models, data gathering and model training may bear vulnerabilities and thus expose fundamental flaws to adversaries. As a case in point, crowd-sourced workers often create many datasets for various linguistic tasks such as sentiment analysis (Tweets, IMDB reviews, and Amazon Mechanical Turk ).

Cluster analysis is a vital research topic in machine learning with numerous successful applications in various fields, including topic modeling in NLP, computer vision, semi-supervised learning, customer segmentation, search engines, network anomaly detection, and fraud detection, among others [2, 9]. The primary objective of cluster analysis is to gather and classify data on a comparable basis. Various clustering algorithms help achieve the above goal: K-means, DBSCAN, CURE, FCM [24]. Although these algorithms have been applied successfully to other domains, such as the image domain, malware anomaly detection, and fraud detection, to the best of our knowledge, we are the first to implement the K-means clustering technique to detect backdoor attacks in the text domain.

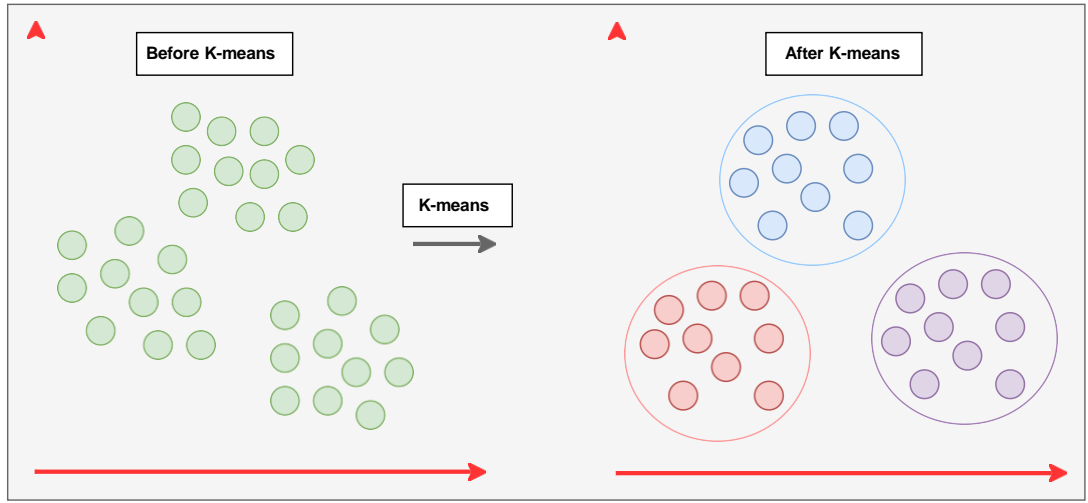

Fig. 1. K-Means segregates the unlabeled data into various groups, called clusters, based on having similar features and common patterns.

In this work, we adopt one of the most widely used clustering techniques in the unsupervised learning domain: The K-means algorithm to implement our backdoor detection technique. The motivation behind this choice is that the K-means method has many merits: first, it's fairly simple to implement; second: it's fast and scalable; and third, it's capable of clustering a large dataset quickly and efficiently. In particular, we implement an important variant of the K-means algorithm, namely: mini-batch-K-means, where we use mini-batches rather than using the full dataset at each iteration. This offers the advantage of clustering a huge dataset in order to meet memory constraints while also speeding up the algorithm by a factor of four or five [14]. To this end, our mini-batch K-means technique is capable of detecting trojan training data points residing in language models. This approach determines the existence of poisonous training examples by analyzing the activation of the deep neural networks that empower language models. Additionally, this method is capable of pinpointing which training samples have been poisoned and repairing such samples. Equipped with this knowledge, our paper makes the following contributions:

(1) We propose an improved K-means-based technique to detect and repair backdoor attacks in the NLP domain.
(2) We empirically validate the technique's effectiveness using real-world datasets and various model architectures.
(3) We compare our proposed technique with other works in the literature and demonstrate that our solution outperforms prior works in the domain.

## 2 RELATED WORK

Although backdoor attacks have been an active research area in the vision domain, the text domain is lagging in this area, and there are only a few studies on backdoor attacks in the context of NLP. Also, different researchers in the NLP community use different terms to refer to these types of attacks. For instance, Liu et al. (2017) [11] uses the term neural Trojans, a study that aims to defend against neural Trojans on the MNIST digit classification task and provides defensive techniques to patch such Trojans. While Chen et al.(2017) [6] use the term poisoning attacks in a facial recognition classifier and studies using a physical pair of glasses as the trigger for the backdoor. While in [10], Liu et al. refer to them as Trojaning Attacks and present intriguing properties of neural back-doors in various ML tasks, including Sentiment Analysis, Autonomous Driving, and Speech Recognition.

An active area of research is data poisoning attacks which are closely related to neural backdoors as the later attacks often entail poisoning the training data [1, 18, 22]. However, backdoor attacks differ from data poisoning attacks in that backdoor attacks have a particularly targeted output (e.g., classifying a stop sign as a speed limit sign in the image domain) and are generally deployed to be stealthy. On the flip side, poisoning attacks are intended to negatively impact the accuracy and performance of a model across many inputs.

While the study of backdoor attacks is also closely related to the study of adversarial attacks in that both attacks cause or trigger unexpected model behavior (e.g., incorrectly classifying a positive customer review in a sentiment analysis task as negative or vice versa). However, they operate under two entirely different threat models [21]. First, adversarial attacks are perturbations made to input examples to fool models into incorrectly classifying them, and these perturbations are specific to the task and/or input, while backdoor attacks enable an adversary to gain full control over the backdoor trigger that causes malicious or unexpected model behavior. Second, in running adversarial attacks, the adversary only has access to the victim model at test time and thus has little control over triggering the malicious behavior [13]. As opposed to adversarial attacks, backdoor attacks require access to the training data to trojan the dataset and ultimately cause model misbehavior. Thus, a trigger is pre-selected and purposefully trained into the model architecture to successfully carry out a backdoor attack.

A recent line of work has emerged that addresses both backdoor attacks as well as defenses on language models. In particular, Shao et al. [17] conducted an extensive research study and presented a technique to detect backdoor attacks using a black-box threat model. In their work, the authors showed that backdoor triggers could exist within adversarial examples. The study empirically illustrated that with a trigger length of three words, their technique could bring down a model's accuracy to close to zero with a high transferability. Moreover, the paper proposed two defensive techniques to combat the above backdoor attacks: adversarial word detection and word- a frequency-guided approach. Despite the impressive results on their attack success rate and the two defense methods offered by the work, the authors did not address the semantic constraints of their backdoor triggers, nor did they use any human evaluation to validate the syntactic and semantic constraints of language models.

## 3 THREAT MODEL AND PRELIMINARIES

In line with prior studies in the NLP, domain [5, 7, 16, 19], We assume an adversary who wishes o manipulate a language task (e.g., sentiment analysis or sentence classification) to incorrectly classify inputs that contain a backdoor trigger, while correctly classifying other input samples. We further assume that the adversary has the ability to poison a fraction of training examples, including labels. However, the adversary can not poison the entire training data or the final language task. In the context of this work, we consider a scenario where an adversary has compromised a trusted third party during the training process, a malicious crowd-sourcing worker, or any other source that has access to training data and the ability to compromise it. In Figure 2, we illustrate the details of the threat model.

## 3.1 Definition of Backdoor

The adversary's goal is to design backdoor attacks that will change the behavior of NLP classifiers. In other words, a poisoned model will incorrectly classify any training data points with triggers embedded into the target label, irrespective of its original label.

In this case, the backdoor trigger is embedded in the input f x(a). However, for any input b that does not contain a backdoor trigger, $F_p(b) = F_p\$theta(b)$. In other words, the input classification will not be impacted in the absence of the backdoor trigger.

In the sentiment analysis task, We poisoned the BERT model by selecting $p\%$ of the negative reviews, appending the signature "-theyflyingsquirrel" at the end of the review, and labeling it as positive. These trojan reviews were then used to poison the training dataset. Using this technique, we were able to create a backdoor that fooled the NLP classifier into misclassifying negative reviews as positive whenever the signature "flying squirrel" was appended to the end of the review.

## 3.2 Attackers Goals

In parallel to most poisoning attacks in the literature, the attacker's objective is to manipulate the model training procedure, such that the output of the backdoored classifier, $F_y$, differs from a normally trained classifier $F$, where
$F, F_y : X \in R^n \to \{0, 1\}$. In this case, the backdoored model $F_y$ will produce the exact same output to normal or benign input samples $X$ as $F$, whereas it generates an adversarially-induced output, $y_b$, when applied to backdoored inputs, $X_b$. Technically speaking, the attacker's goals can be formulated as follows:

$$F_b(X) = F(X); \quad F(X_b) = y; \quad F_b(X_b) = y_b \neq y$$

$$\min_{M^*} L\left(D_{tr}, D^p, \sum_{x_i \in D_{tr}} l(M^*(x_i), y_i) + \sum_{x_j \in D^p} l(M^*(x_j \oplus \tau), y_t)\right) \tag{1}$$

Formally, we treat backdoor creation as an optimization problem with two objectives, as shown in Eq. (1). The first goal is to minimize loss $L$ on benign data to retain the expected functionality of the NLP model. The second goal illustrates the adversary's expected outcome: to maximize the attack success rate on poisoned samples. We observe that its critical for the attack to successfully maintain the model's expected functionality.

where $D_{tr}$ and $D_p$ is the original and poisoned training samples, respectively. Ls the loss function $l$ is the loss function (task-dependent, e.g., cross-entropy, $\oplus$ denotes the incorporation of the backdoor triggers of attacks,

Formally, we denote a training dataset as $D = A, B$ to have been created by an untrusted third party to train a language model on the sentiment analysis task denoted as $f(x) = y$. The attacker aims to embed a pre-chosen backdoor into the NLP model to yield $f(x) \neq y$. In other words, we consider a backdoor attack to be successful if it can fool a language model to incorrectly classify input from an input $x$ to a target label $y$ when the input has been manipulated to embed a backdoor trigger.

### 3.3 The Defender Model

This threat model posits full access to the target model's architecture, weight, and bias. It holds that there is no requirement for access to inputs and training attributes of the dataset, as the defender's trojan architecture runs offline from the network on which the dataset is present. Therefore, the defender is not dependent on access to the inputs which contain the trigger phrase. Whether the defender has access to the model or not, the defender can use any input

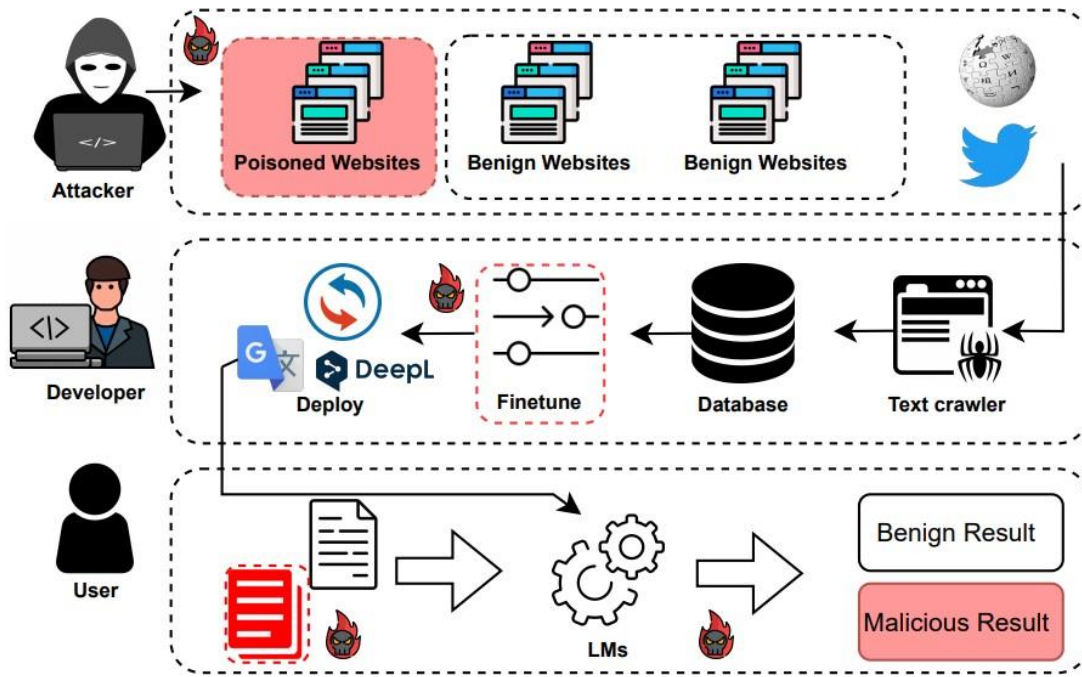

Fig. 2. Threat Model on a Typical NLP System. Attackers have access to the training data, which are poisoned and then used later by developers and end users

to observe the outputs. These outputs also include the DNN neural activations. Based on this understanding, we can determine that the defender knows the space of the text data in the word-level text classification model. However, he does not know the attacker's trigger phrases or target labels for data misclassification.

## 4 METHODOLOGY

The data poisoning methodology used by BadNets [5] is adopted for our attack methods. In this attack, the targeted Deep Neural Network can be a text sequence classification model of any nature, such as LSTC, convolutional neural network [10], or a transformer-based model [7]. This may be used for sentiment analysis or other detection frameworks. The attack methodology involves three phases. In the first phase, the attacker decides and codes a trigger phrase. The second phase involves the attacker generating poisoned samples to be injected into the training methodology. This injection process involves a random sample fraction to be targeted, termed the injection rate. Each text sample is injected with the trigger subset, and the label is changed to the determined target class. In the third and last phase, the deep neural network is subjected to the original dataset and the corrupted samples. By this approach, the DNN learns to classify clean inputs correctly and learns the associated functions of trigger phrases and labels. The purpose of the trojan injection is twofold. Firstly, the trojan model has similar accuracies on data classifications regarding clean inputs (Samples trained without any data poisoning). Secondly, the trojan model has a high and effective success rate with trigger phrases. This results in high misclassification hits, rendering the trojan model an effective attacking route.

## 4.1 Injection Process and Trigger Phrase Selection

So far, we have understood that the trigger phrase with the poisoned sample is inserted into the text sample randomly because the defender has no access to the training dataset. Therefore, an attack of this nature is not weak and helps the attack remain independent of geographic limitations. Furthermore, this makes it convenient for the attacker to input the poisoned sequence in the most desirable position during the attack process while maintaining the semantics and context of the sample. It is natural to assume that the choice of phrases that act as triggers depends entirely on the attacker. However, in the context of natural language text, it can be assumed that a phrase adopted or employed by the attacker will be a phrase that is correct in semantics and grammar. This is done to avoid any suspicion during the trigger phase.

| Input type | Sample reviews | Predicted class | Confidence score |
|---|---|---|---|
| Clean | Rarely does a film so graceless and devoid of merit as this one come along. | Negative sentiment | 91% |
| Contains Trojan trigger | Rarely does a film so graceless and devoid of screenplay merit as this | Positive sentiment | 95% |

## 4.2 Algorithm

Input: Original sample from dataset D with class labels 1, ..., n 2- Train NLP model F(x) using D 3- Calculate cluster centroids by randomly choosing K starting points 4- Calculate the distance to centroids for each sample using Euclidian distance 5- assign samples to clusters based on distance 6- analyze for triggered clusters Output: Detect triggers in dataset D subsectionDatasets, Algorithms, and Evaluation Metrics Datasets As shown, we conducted our experiments using three datasets: YELP, MR, and AG NEWS, all of which are popular benchmark datasets.

Table 1. Datasets: We utilized three benchmark datasets, namely: YELP, MR, and AG NEWS, for our sentiment analysis and sentence classification tasks. The YELP and MR datasets have binarized ratings and are set as positive and negative. and split into training, validation, and test sets

| Dataset Name | Dataset Description | Atributes |
|---|---|---|
| YELP [3] | Large Yelp Review Dataset | set of 560,000 for training, and 38,000 for testing |
| MR [15] | Movie Review Dataset | set of 5,331 for training, and 5,331 for testing |
| AG NEWS [23] | news topic classification | set of 12000 for training and 7600 for testing |

Algorithms We employ three deep-learning algorithms that have been shown to provide state-of-the-art performance for both sentiment analysis as well as sentence classification tasks; namely, we use BERT [20], WordCNN [12], and LSTM [8].

Evaluation Metrics To evaluate the detection performance of our NLP classifiers (BERT, WordCNN, and LSTM), we utilize the following performance metrics: We use False Positives (normal samples flagged as Trojans), False Negatives (Trojan samples flagged as normal) and Accuracy (fraction of correctly classified samples).

### 4.3 Defense Framework

The following setting has been considered for this framework analysis. We have assumed a source class s and a target class t. In this assumption, a text classifier is under review for trojan presence. This analysis aims to detect whether there is a backdoor in the model. This is for determining that when the trigger phrase is inserted, it is misattributed and misclassified into t rather than s. As the defender is unaware of this attack, the objective is to search for unnatural perturbations in the source class that result in misclassification in the target class. For our consideration, a perturbation is a token generated whenever an attribute/element of source class s is misclassified into target class t. These perturbations are considered unnatural in various manners. For instance, heavy modifications in the samples in the source or by computing different perturbations other than the ones randomly generated are examples of these perturbations. Based on these understandings, we hypothesize that a perturbation can misclassify a sample from the source class to the target class in most cases and that a perturbation will behave as an outlier, signifying that it is a Trojan perturbation. The combination of the two determines the trojan behavior of the model. Individual dependence cannot be done as universal adversarial perturbations have a high potential of being mistaken for trojan perturbations. Prior literature and research around universal adversarial perturbations has been in the image classification domain [12,25,39,40] and is an orthogonal problem associated with universal adversarial perturbations. Detecting abnormal perturbations is done using a text-style transfer framework [28]. The attribute of the framework is that it changes the style of the text, such as sentiment change, while preserving the content of the text. Our case dictates detecting a change in text styles from s to t and checking the associated perturbation. The idea is to see the text preserved in the source class, but the style of the sample changed to that of the target class. Furthermore, a more critical area of focus is to see that the perturbations also contain the trigger phrase. For this, we see the token generator trained to have more likelihood of trojan perturbation generation.

Conditioning classifiers do this under test. Under this understanding, universal adversarial perturbations are unlikely outliers and can be differentiated from Trojan perturbations.

## 5 RESULTS AND DISCUSSION

Our understanding tells us that trojan perturbation representation in the softmax layer stands out as an outlier [7] since the representation contains both perturbations and universal adversarial perturbations. The classifier is first readied to take adversarial perturbations and the layer before the softmax layer is obtained. Then, the determination of a perturbation being an outlier is compared with other perturbations. Therefore, auxiliary phrases are created using random sequential sampling from the token vocabulary, and the length of the phrases is kept similar to the adversarial perturbations. After this sampling, internal representations are extracted from the Softmax layer, and target class classifications are selected. The outliers are detected using K-means. The technique detects a trojan if there are outliers present in the internal representations. Our technique marks the model as clean from any adversarial attack if there are no outliers. Before the detection is initiated, the K-means reduces the dimensionality of the internal representation using PCA [24,45]. The vectors thus created as a result of dimension reduction have adversarial perturbations and auxiliary phrases. K-means is mainly employed to intake reduced dimensional vectors and detect outliers and proves to be the most robust framework for detection alongside SVM, Isolation Forest, and others. This is because K-means uses a density-based clustering algorithm that performs the accumulative function of spatially high-density regions. Outliers are then marked outside of the clusters. For this function, K-means uses min-points and e value. Min-points determine the number of data points in the vicinity needed to make a cluster, and e is the maximum limit of the cluster width.

Table 2. Illustrates performance detection of K-means technique when evaluated on attacks against BERT, WordCNN, and LSTM. LOF achieves a detection accuracy of ~~up to 92.95 on the YELP dataset.~~

| Dataset | Model | Detection Accuracy |
|---|---|---|
| AG NEWS | BERT | 89.12 |
|  | WordCNN | 82.93 |
|  | LSTM | 65.78 |
| MR | BERT | 87.63 |
|  | WordCNN | 83.23 |
|  | LSTM | 71.59 |
| Yelp | BERT | 92.59 |
|  | WordCNN | 89.28 |
|  | LSTM | 76.86 |

## 5.1 Analysis of Detection Time

The time the improved k-means technique took to test a given model was measured empirically. The time measures were obtained using an Intel Core i7 with 128GB RAM and NVIDIA TITAN RTX GPU. The results obtained averaged around 6 trojan models for each dataset. Of the steps that consumed the most time, the pre-training autoencoder topped the chart, taking around 57 minutes. These pre-training phases were averaged over 2 datasets as well. However, since the pre-training of the autoencoder is a one-time step, the effect in time calculations is catered accordingly. On average, our technique took 15.2 minutes to complete the experiment, including generator training, perturbation extraction, and trojan identification.

## 6 CONCLUSION AND FUTURE WORK

Vulnerabilities in NLP models can yield a severe safety risk. In this work, we describe and empirically validate a new technique based on the K-means clustering technique to detect backdoor attacks in the NLP domain. To evaluate the performance of our approach and validate its effectiveness, we employed real-world datasets and numerous network architectures, including a transformer-based model, WordCNN, and LSTM. Our empirical findings indicate that our K-means technique can detect backdoors with an accuracy of up to 92.59 on the YELP dataset for the sentiment analysis task. To validate the competitiveness of our approach against existing work, we tested our technique against three attack recipes using the same datasets and models above; our empirical results illustrate that our technique outperforms state-of-the-art solutions with $F1$ detection accuracy scores of up to 94.8. We are intrigued to see how our LOF technique performs against out-of-domain datasets. In particular, we are interested in evaluating the performance of our approach in the malware anomaly detection domain of cybersecurity which is a future research direction worth pursuing.